\documentclass[twocolumn,showpacs,preprintnumbers,amsmath,amssymb]{revtex4}

\draft % marks overfull lines with a black rule on the right

\usepackage{graphicx}% Include figure files
\usepackage{dcolumn}% Align table columns on decimal point
\usepackage{bm}% bold math
\usepackage{epstopdf}
\begin{document}

%\preprint{AIP/123-QED}

\title{Dry-Mass Sensing for Microfluidics}% Force line breaks with \\

\author{T. M\"uller}
\author{D. A. White}
\author{T. P. J. Knowles}
\email{tpjk2@cam.ac.uk}
\affiliation{Department of Chemistry, University of
Cambridge, Lensfield Road, Cambridge CB2 1EW, United Kingdom}
\date{\today}% It is always \today, today,

\begin{abstract}
We present an approach for interfacing an electromechanical sensor with a microfluidic device for the accurate quantification of the dry mass of analytes within
microchannels. We show that depositing solutes onto the active
surface of a quartz crystal microbalance by means of an on-chip microfluidic
spray nozzle and subsequent solvent removal provides the basis for the
real-time determination of dry solute mass. Moreover, this
detection scheme does not suffer from the decrease in the sensor's quality factor and the viscous drag present
if the measurement is performed in a liquid environment, yet allows
solutions to be analysed. We demonstrate the sensitivity and
reliability of our approach by controlled deposition of nanogram
levels of salt and protein from a micrometer-sized channel.
\end{abstract}

%Use showkeys class option if keyword
\maketitle

Label-free, quantitative detection of small quantities of
biomolecules underlies experimental approaches in a wide range of fields
spanning industry and applied as well as basic research in numerous
fields ranging from physics and chemistry to biology, food science
and medicine \cite{Jones1996,Fink1998,Chiti2006,Ubbink2008,Aguzzi2010}. In
such applications, microfluidic strategies \cite{Whitesides2006,Baker2009,Wu2011}
allow for an enhanced level of control over the specific environment \cite{Fodera2012} and offer a wide range of
preparation and separation
techniques \cite{Brody1997,Ahn2006,Pamme2007,Lenshof2010,Abate2010a}. The
ability for accurate label-free measurements in micro
technology platforms opens up fruitful possibilities for 
multidisciplinary research and has the propensity to advance our
understanding of biology as well as offer exciting perspectives for the
development of bio-inspired nanomaterials \cite{Fowler2006,Fowler2007,Otzen2010,Aizenberg2010,Knowles2011b,Hol2014}.

To fulfill the demands of the growing complexity of these analyte
systems, there is currently a pressing need for multi-purpose,
high-sensitivity measurement strategies. We approach this challenge
by combining the versatility of microfluidics with the precision of
electromechanical sensors (EMS), without the impediment of the
substantial decrease of the resonator quality factor in liquid.
Utilising EMS as detectors is highly attractive in terms of ease of
use and cost. Furthermore, while electromechanical transducers do not provide any information
directly about the charge-to-size ratio, which is available from
mass-spectrometry \cite{Sikanen2010}, they are independent of
ionisation energies and can therefore yield quantitative results on
the total mass. Moreover, interfacing with microfluidic continuous flow
separation techniques \cite{Pamme2007,Lenshof2010} in principle could
allow for determination of a large variety of physical properties of
biomolecules in solution such as size or charge.

\begin{figure}[ht]
\centering
\includegraphics{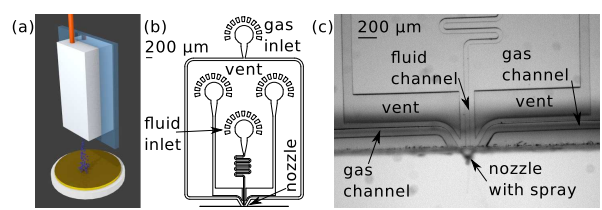}
\caption{\label{fig:Pics} (a) Schematic representation of the
measurement setup consisting of a microfluidic device spraying a
controlled volume of a solution onto the gold electrode of a quartz
crystal microbalance. (b) Design of a microfluidic nozzle. (c)
Micrograph of a spraying device with a continuous flow rate set to
1~ml/h and an exposure time of $30~\mu\textrm{s}$. The scale bars
are $200~\mu\textrm{m}$.}
\end{figure}
%\begin{figure*}[ht]
%\centering
%\includegraphics[width=0.8\textwidth]{Pics.pdf}
%\caption{\label{fig:Pics} (a) Schematic representation of the measurement setup consisting of a microfluidic device spraying its contents onto the gold electrode of a quartz crystal microbalance.  (b) Design of a microfluidic nozzle. (c) Micrograph of a spraying device with a continuous flow rate set to 1~ml/h and an exposure time of $30~\mu\textrm{s}$. The scale bars are $200~\mu\textrm{m}$.}
%\end{figure*}

More specifically, we have developed an ex-situ detection method consisting of a microfluidic chip
interfaced with a quartz crystal microbalance (QCM) measuring the
dry mass of the microchannel's contents. In practice, the analyte
within the chip is deposited onto the active surface of the
microelectromechanical sensor through a microfluidic
nozzle \cite{Thiele2011,Amstad2012} driven by pressurised gas
co-flowing with the liquid at end of the channel \cite{Muller2013}.
This scheme is depicted in Fig.~\ref{fig:Pics}(a) and a micrograph
of the spray nozzle in action is shown in Fig.~\ref{fig:Pics}(b).

To allow for quantitative analysis of the solute inside the channel,
the solvents impinging on the microbalance need to be evaporated
completely. This objective has been effected in macroscopic
approaches by stopping flow and heating the sensor
electrodes \cite{Schulz1973} or by immersing the analyte in
supercritical fluids \cite{DiMilia1992}. In our integrated
microscopic strategy, a mechanical shutter prevents deposition while
it is closed, permitting \textit{continuous} operation of the
microfluidic device resulting in very accurate flow from the nozzle.
It should be noted that our approach effectively decouples the
processing and measurement stages, and thus benefits from the
optimal performance of both integral parts. In particular,
combination with microfluidic separative
methods - such as diffusive or
electrophoretic migration - in the incorporated microfluidic device
would provide specificity to our detection module, creating a simple, highly sensitive and versatile microanalytical
platform in the shape of a ``lab on two chips''. In such an approach, separation
occurs in solution whereas the measurement is performed in air,
significantly enhancing applicability and sensitivity as the quality
factor of the resonator in not adversely affected by viscous drag.

Quartz crystal microbalances are an established technique for the
determination of mass \cite{Sauerbrey1959,Hlavay1977}, and there is
also a variety of electro- and purely mechanical sensors operating
in a macroscopic liquid
environment \cite{Kanazawa1985,Rickert1997,Welland2007,Hovgaard2007,Knowles2007a,White2009}
as well as \emph{inside} microfluidic
devices \cite{Godber2007,Doy2010,Hill2010,Yu2010,Hu2012} or containing microfluidic channels
themselves \cite{Burg2003,Verbridge2005}. While these latter
approaches yield good results in terms of sensitivity, the operation
of QCMs in liquid suffers from the reduction in their quality factor
due to viscous damping and quantitative measurement of (dry) mass
remains challenging. In our ``lab-on-two-chips'' approach the
automated two-step operation - deposition of analyte and subsequent
evaporation of solvents utilising a mechanical shutter - allows for
simple, continuous operation and enables us to straightforwardly
access the analyte's dry mass.

%\section*{Materials and Methods}

Microfluidic devices were fabricated to a height of
$25~\mu\textrm{m}$ using standard soft lithography techniques in
polydimethylsiloxane (PDMS, Sylgard 184, Dow Corning, Midland MI,
USA) on SU8 masters \cite{Duffy1998}. As can be seen in
Fig.~\ref{fig:Pics}(b) the device contains a $10~\mu\textrm{m}$ wide
horizontal line at the nozzle ensuring reproducible fabrication.
This template line can be used to position a razor blade to cut the
device at the exact position of the nozzle. Upon plasma activation
of the PDMS devices and the microscope glass slides, the edge of the
glass slide can be aligned to the edge of the device by pressing the
side of a razor blade (or similarly flat object) laterally to the
pdms and placing the glass slide on top, also pressing it against
the blade.

\begin{figure}[ht]
\centering
\includegraphics{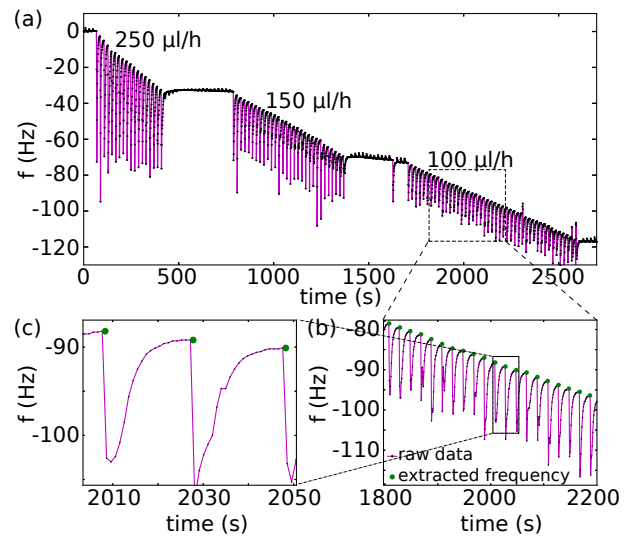}
\caption{\label{fig:concept} (a) Measured frequency shift as a
function of time for the deposition of a 10 mM NaCl solution at
three different flow rates. A shutter controls spraying during 0.5~s
followed by 19.5~s for drying and equilibration, leading to the
sharp flanks and subsequent settling of the resonance frequency. (b)
Magnification of a part of (a) with the flow rate set to
$100~\mu\textrm{l/h}$. The green dots represent the maximum
frequency within each 20~s interval. (c) Zoom on individual spray
bursts from (a) and (b).}
\end{figure}

Using a syringe pump (Harvard Apparatus PHD2000) and precision glass
syringes (Hamilton Bonaduz gastight 1700 and 1800 series with
volumes between 50 and $1000~\mu\textrm{l}$) connected via
polyethylene tubing (Smiths Medical, 800/100/120), a controlled flow
of liquid is driven through the fluid inlet (marked in Fig.~\ref{fig:concept}(b)) of a microfluidic channel. This channel is
open at the edge of the glass slide where it is met by two further
channels at an angle of 60 degrees (see Figs.~\ref{fig:Pics}(b) and
(c)) through which pressurised nitrogen gas at ca.~4~bar is
applied, forming a microfluidic spray nozzle. We found that the reliability of continuous operation of this design could be enhanced considerably by fabricating on the same chip a set of vented
channels between the liquid and gas channels to prevent
the gas from being transported into the fluidic channel due to the
gas-permeable nature of PDMS.

The solute is deposited on the microbalance's active surface through
a microfluidic nozzle at which a spray is formed via pressurised
gas \cite{Thiele2011}. A schematic representation thereof is shown in
Fig.~\ref{fig:Pics}(a). Alternatively, an electrospray could be
generated \cite{Ramsey1997} or a glass capillary could be attached to
the analysis outlet of an existing device, at the end of which a gas flow creates a spray.
The latter approach would allow for straightforward incorporation
into multifunctional microfluidic systems.

To obtain absolute values of the mass deposited in this manner, a
calibration measurement can be carried out by spraying ions
of a known concentration onto the balance and determining the
ensuing shift in resonance frequency.

In order to evaporate all liquid from the spray and to equilibrate
the QCM, a shutter controlled with a stepper motor is closed after
0.5~seconds of spraying for 19.5~s. Furthermore, to protect the
sensitive apparatus from fast temperature fluctuations and air
currents the microbalance and the microfluidic chip were placed in
an incubator casing (with the temperature control turned off).
Alignment between the microbalance and the chip is achieved by a
``helping hand'' clamp holder which allows for simple positioning of the microfluidic chip with respect to the QCM. However, incorporation of a $xyz$-micrometer stage with coordinates relative to the centre of the quartz chip is readily achievable.

Readout of the QCM crystal (Stanford Research Systems 100RX1, Cr/Au, 5~MHz) is performed
with a commercial frequency counter (Stanford Research Systems QCM200) at a gate time of
1~s, leading to an accuracy of 0.1~Hz. The sensor has an electrode
surface of 1.37~cm$^2$, with its active area confined to
0.40~cm$^2$ by the geometry of the second electrode, and a mass
sensitivity coefficient of 0.0566~Hz/ng.

After each experimental run, the microbalance was unmounted and rinsed with deionised water to remove adsorbed solutes which recovered its original properties.

%\section*{Results and Discussion}

In Fig.~\ref{fig:concept}(a) the functionality and reliability of
this approach is demonstrated. Shown is the frequency shift of a
5~MHz quartz crystal microbalance upon the controlled deposition of
10~mM NaCl solution. By initiating the flow of the solution and
opening the shutter for a time period of $t=500~\textrm{ms}$, a
pre-defined volume of solution is sprayed onto the balance. This
deposited mass is determined by $m=Q\cdot t\cdot M$, with $Q$
being the volume flow rate and $M$ being the molar mass of the
compound, and results in a downward spike in the reading of the
QCM's resonance frequency. After all liquid has evaporated, the
frequency reading saturates at a value shifted with respect to the
original one corresponding to the mass of the solute deposited. Flow
rates of 250, 150 and $100~\mu\textrm{l/h}$ were applied, yielding
volumes of 35, 21 and 14~nl or masses of 20, 12 and 8~ng,
respectively, per 0.5~s spray burst - well below the limit of micropipettes.

Our results show that the equilibrated values of the frequency shifts - resulting from repeated cycles of solution deposition and solvent removal - display a remarkably
linear relationship, the slope of which is proportional to the flow rate and stable over
tens of minutes. After the flow is stopped, a well-defined base-line
is rapidly attained after a short settling time of the order of a
minute due to the change in temperature caused by the ceased
evaporative cooling. These measurements demonstrate a dynamic range
of over three orders of magnitude for the detected mass, from nanograms to micrograms.
The linear resonse is maintained up to a point where several micrograms of solutes have been deposited during continuous operation over time scales exceeding thousands of seconds. In a small fraction of cases, an individual spray burst was observed to result in a transient peak that increased rather than decreased the resonance frequency. However, upon evaporation the overall negative frequency shift was restored. Such spikes can be removed readily with a suitable
data-processing algorithm which is checking the direction of the
frequency shift upon deposition.

A detailed analysis of several consecutive spray bursts
(Figs.~\ref{fig:concept}(b) and (c)) shows even more clearly the
precision obtained with the presented approach. Here, the green dots
are the extracted frequency reading obtained by taking the maximum
frequency within a 20~s interval. Readout noise could be further
reduced by synchronising the frequency measurement with the shutter
opening or using a flank-detection algorithm and averaging the last
few frequency points before the shutter is reopened.

\begin{figure}[ht]
\centering
\includegraphics{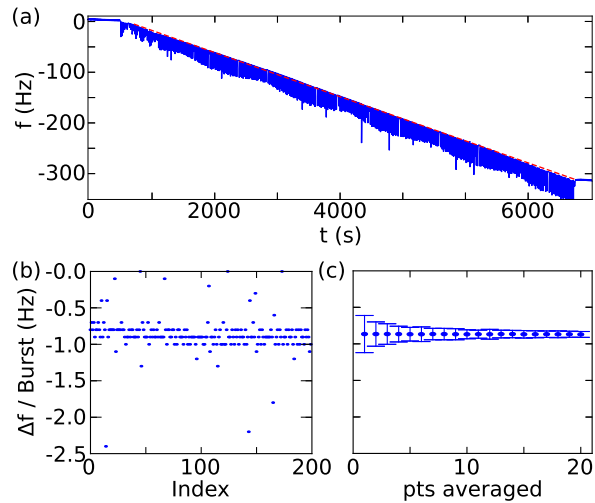}
\caption{\label{fig:avg} (a) Spray deposition of lysozyme at a
concentration of 1~mg/ml over a time scale of 2 hours with a shutter
opening time of 0.5~s every 20~s and a flow rate of
$100~\mu\textrm{l/h}$. The dashed red line is a guide to the eye.
(b) Scatter plot of the frequency shift of the central 200
individual burst in (a). Note that the detector circuit's gate time
of 1~s leads a frequency resolution of 0.1~Hz. (c) Average frequency
shift and corresponding standard deviation per burst when averaging
over consecutive points.}
\end{figure}

To illustrate further the precision of this method, we have recorded
the frequency shift due to the deposition of 1~mg/ml of the model protein lysozyme (without NaCl) at a flow rate of $100~\mu\textrm{l/h}$ - i.e., continuously spraying
$100~\mu\textrm{g/h}$, $2.5~\mu\textrm{g/h}$ of which are deposited
on the QCM - over two hours (see Fig.~\ref{fig:avg}(a)). A linear
response is conserved during the entire measurement period of two
hours. In Fig.~\ref{fig:avg}(b) the resulting frequency shift of 200
individual spray burst is shown. It can be seen that most of the
burst result in a shift of between -0.7 to -1.0~Hz, with minute
drift of around -0.05~Hz/h and just a dozen points above and below
this window in a range of 0 to -2.4~Hz. These outliers might result
from ambient influences that can interfere with the spray
deposition, inaccuracies in the shutter opening time, or
combinations thereof. Averaging the frequency shifts of a small
number of consecutive bursts - as presented in Fig.~\ref{fig:avg}(c)
- naturally decreases the standard deviation of this average signal,
with values ranging from $\pm0.25~\textrm{Hz}$ without averaging
over $\pm0.09~\textrm{Hz}$ for 5 points to $\pm0.04~\textrm{Hz}$ for
20 averages. For the remainder of this work we have chosen to
average over 5 points as a compromise between small amount of mass
per averaged value and superior signal-to-noise ratio.

\begin{figure}[ht]
\centering
\includegraphics{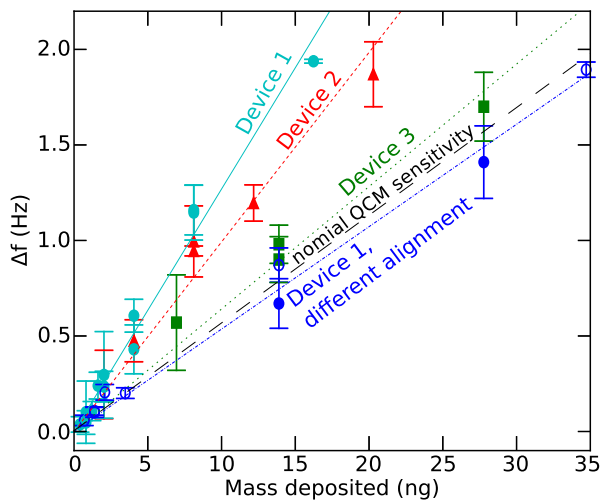}
\caption{\label{fig:deposition} Measured frequency shift per 0.5~s
spray burst averaged over 5 consecutive bursts as a function of
actual mass deposited. The data are for 3 different devices (device
1: circles, device 2: triangles, device 3: squares) with different
alignments represented by different colours and using predetermined
concentrations of NaCl (filled symbols) and lysozyme (hollow
symbols). The lines are fits to each of the individual data sets.}
\end{figure}

To benchmark the performance of this technique the frequency shift
resulting from a 500~ms spray burst is presented as a function of deposited mass in
Fig.~\ref{fig:deposition}. The sprayed mass has been calculated from
the volume of liquid flowing through the device in 0.5~s and the
pre-set concentration of analyte ranging from 1 to 10~mM and 1~mg/ml
NaCl (filled symbols) as well as 0.1 and 1~mg/ml lysozyme (without any salt; empty
symbols). The frequency values shown are the mean and
standard deviations taken from tens to hundreds of spray bursts
averaged over five consecutive shutter openings.

The data shown are taken with 3 nominally identical devices
(different symbols), with one device studied at two different
relative alignments of the nozzle with respect to the QCM (different colours), resulting in 4 distinct
data sets. We note that for each data set the different flow rates
and analyte concentrations yield frequency shifts that are linearly
proportional to the mass deposited. Thus we are able to first
calibrate our measurement setup using a well-defined analyte
concentration and consequently measure unknown amounts of solute to
a high degree of accuracy. It is thus possible to achieve
nanogram precision using minute sample quantities. The slopes can
vary up to roughly a factor of two between experimental runs due to
the alignment of spray nozzle and microbalance in combination with
the non-uniform mass sensitivity of the QCM. In particular, a
well-aligned device - where the spray hits the microbalance
centrally - can exceed the average nominal sensitivity of the
QCM (widely-dashed black line). By measuring accurately the position
of the spray deposited, the expected frequency shift per unit mass could in principle be
calculated precisely \cite{Cumpson1990}. Since we can
calibrate the mass sensitivity by running a
solution with well-known concentration first, such a cumbersome
process is, in fact, unnecessary.

%\section*{Conclusion and Outlook}
Our measurements show that 5 bursts of as little as a few nanograms
of deposited mass each are enough for the precise determination of the
analyte concentration within a microfluidic channel. Thus, we have
demonstrated an accurate and quantitative label-free detection
method for microfluidic technologies by exploiting the high
sensitivity of a commercially available quartz crystal microbalance
(0.0566~Hz/ng).

Traditional microfluidic methods work with the
microelectromechanical sensor placed in liquid or at a liquid-air
interface. Therefore these techniques are heavily affected by
viscous drag of particles in liquid, and consequently they are not
able to quantitatively resolve the dry mass of the analyte.
Moreover, these strategies require the surface of the sensor to be
functionalised in order to guarantee adherence of the sample
molecules \cite{Spangler1999,Nicu2008}. The approach discussed here circumvents both of these difficulties by transporting the analyte out of the channel and onto the sensor, where it is
dried. Once all solution has evaporated - a process that is
automatically taking place while a mechanical shutter decouples the
continuous spray from the sensor - gravitation and surface adhesion
hold all solutes in place. Nevertheless, if desired, specificity could in principle be recovered at the price of generality by functionalising the QCM surface in combination with a washing step after every spray burst.

The detection method described in the present paper provides a highly accurate and versatile tool for quantitative assays. Measurements of analyte mass have proved to be highly successful in other areas of analytical chemistry, and the integration between EMS and microfluidics opens up the possibility of quantitative mass measurements in a format that is compatible with conventional PDMS microfluidics. Crucially, the approach for mass measurement discussed in the present work is not dependent on the surface chemistry of the sensor nor on specific binding events, requirements that underlie conventional use of QCMs in liquid. This approach therefore represents a label-free detection module that is applicable to volumes and concentrations that are of interest to lab on a chip applications. In order to fully harness the potential of this measurement strategy using complex analytes - such as biological samples - the detection module could be coupled to microfluidic upstream separation techniques such as diffusive filtering\cite{Brody1997} or free-flow electrophoresis \cite{Pamme2007,Lenshof2010}.

We thank Alexander K. Buell, Igor Efimov, and Victor Ostanin for
valuable discussions on QCM sensors and gratefully acknowledge
financial support from the Swiss National Science Foundation (SNF),
the Engineering and Physical Sciences Research Council (EPSRC),
Biotechnology and Biological Sciences Research Council (BBSRC), the
European Research Council (ERC), as well as the Frances and Augustus
Newman Foundation.

%\section*{References}
%\bibliography{/home/mueller/Publications/Bibliography} %your .bib file
%\bibliographystyle{style}

\end{document}